\documentstyle[psfig]{mn}

\newif\ifAMStwofonts

\ifoldfss
  \ifCUPmtlplainloaded \else
    \NewTextAlphabet{textbfit} {cmbxti10} {}
    \NewTextAlphabet{textbfss} {cmssbx10} {}
    \NewMathAlphabet{mathbfit} {cmbxti10} {} 
    \NewMathAlphabet{mathbfss} {cmssbx10} {} 
  \fi
  \ifAMStwofonts
    \ifCUPmtlplainloaded \else
      \NewSymbolFont{upmath} {eurm10}
      \NewSymbolFont{AMSa} {msam10}
      \NewMathSymbol{\upi}     {0}{upmath}{19}
      \NewMathSymbol{\umu}     {0}{upmath}{16}
      \NewMathSymbol{\upartial}{0}{upmath}{40}
      \NewMathSymbol{\leqslant}{3}{AMSa}{36}
      \NewMathSymbol{\geqslant}{3}{AMSa}{3E}

    \fi
  \fi
\fi 

\ifnfssone
  \newmathalphabet{\mathit}
  \addtoversion{normal}{\mathit}{cmr}{m}{it}
  \addtoversion{bold}{\mathit}{cmr}{bx}{it}
  \newmathalphabet{\mathbfit} 
  \addtoversion{normal}{\mathbfit}{cmr}{bx}{it}
  \addtoversion{bold}{\mathbfit}{cmr}{bx}{it}
  \newmathalphabet{\mathbfss} 
  \addtoversion{normal}{\mathbfss}{cmss}{bx}{n}
  \addtoversion{bold}{\mathbfss}{cmss}{bx}{n}
  \ifAMStwofonts
    \ifCUPmtlplainloaded \else
      \UseAMStwoboldmath
      \makeatletter
      \new@mathgroup\upmath@group
      \define@mathgroup\mv@normal\upmath@group{eur}{m}{n}
      \define@mathgroup\mv@bold\upmath@group{eur}{b}{n}
      \edef\UPM{\hexnumber\upmath@group}
      \new@mathgroup\amsa@group
      \define@mathgroup\mv@normal\amsa@group{msa}{m}{n}
      \define@mathgroup\mv@bold\amsa@group{msa}{m}{n}
      \edef\AMSa{\hexnumber\amsa@group}
      \makeatother
      \mathchardef\upi="0\UPM19
      \mathchardef\umu="0\UPM16
      \mathchardef\upartial="0\UPM40
      \mathchardef\leqslant="3\AMSa36
      \mathchardef\geqslant="3\AMSa3E
    \fi
  \fi
\fi 

\ifnfsstwo
  \DeclareMathAlphabet{\mathbfit}{OT1}{cmr}{bx}{it}
  \SetMathAlphabet\mathbfit{bold}{OT1}{cmr}{bx}{it}
  \DeclareMathAlphabet{\mathbfss}{OT1}{cmss}{bx}{n}
  \SetMathAlphabet\mathbfss{bold}{OT1}{cmss}{bx}{n}
  \ifAMStwofonts
    \ifCUPmtlplainloaded \else
      \DeclareSymbolFont{UPM}{U}{eur}{m}{n}
      \SetSymbolFont{UPM}{bold}{U}{eur}{b}{n}
      \DeclareSymbolFont{AMSa}{U}{msa}{m}{n}
      \DeclareMathSymbol{\upi}{0}{UPM}{"19}
      \DeclareMathSymbol{\umu}{0}{UPM}{"16}
      \DeclareMathSymbol{\upartial}{0}{UPM}{"40}
      \DeclareMathSymbol{\leqslant}{3}{AMSa}{"36}
      \DeclareMathSymbol{\geqslant}{3}{AMSa}{"3E}
    \fi
  \fi
\fi 

\ifCUPmtlplainloaded \else
  \ifAMStwofonts \else 
    \def\upi{\pi}
    \def\umu{\mu}
    \def\upartial{\partial}
  \fi
\fi

\title{Precision Timing Measurements of PSR~J1012+5307}
\author[Lange et al.]{ Ch.~Lange,$^1$
F. Camilo,$^{2}$ N. Wex,$^1$ M. Kramer,$^{3}$ D.C. Backer,$^4$
A.G. Lyne,$^{3}$ \newauthor
O. Doroshenko$^{1}$ \\
$^1$Max Planck Institut f\"ur Radioastronomie, Auf dem H\"ugel 69, 
D-53121 Bonn, Germany \\
$^2$Columbia Astrophysics Laboratory, Columbia University, 550 West 120th
Street, New York, NY 10027, USA \\
$^3$University of Manchester, Jodrell Bank Observatory, Macclesfield,
Cheshire SK11 9DL, UK \\
$^4$Astronomy Department, 601 Campbell Hall, University of California,
Berkeley, California 94720-3411, USA}

\date{21 January 2001}

\pagerange{\pageref{firstpage}--\pageref{lastpage}}
\pubyear{2001}

\begin{document}

\maketitle

\label{firstpage}

\begin{abstract}
We present results and applications of high precision timing
measurements of the binary millisecond pulsar J1012+5307.  Combining
our radio timing measurements with results based on optical
observations, we derive complete 3-D velocity information for this
system. Correcting for Doppler effects, we derive the intrinsic spin
parameters of this pulsar and a characteristic age of $8.6\pm1.9$
Gyr. Our upper limit for the orbital eccentricity of only
$8\times 10^{-7}$ (68\% C.L.) is the smallest ever measured for a
binary system. We demonstrate that this makes the pulsar an ideal
laboratory to test certain aspects of alternative theories of
gravitation. Our precision measurements suggest
deviations from a simple pulsar spin-down timing model, which are
consistent with timing noise and the extrapolation of the known
behaviour of slowly rotating
pulsars. 
\end{abstract}


\section{Introduction}
\label{intro}

The 5.3-ms pulsar J1012+5307 was discovered during a survey for short period
pulsars with the 76-m Lovell radio telescope at Jodrell Bank.  Nicastro et
al.~\shortcite{nll95} showed that the pulsar is in a binary system with an
orbital period of 14.5 hr and a companion mass between 0.11\,M$_{\odot }$ and
0.25\,M$_{\odot }$ (90\% confidence level, C.L.). 
Optical observations reported by Lorimer et al.\ \shortcite{llf95} 
revealed an optical counterpart within 
0\farcs 2$\pm$0\farcs 5 of the pulsar timing position, 
consistent with its being a white dwarf (WD) companion of 
mass 0.15\,M$_{\odot }$.

Indeed, PSR~J1012+5307 is one of the few examples of binary millisecond
pulsars where optical observations of the WD companion considerably enhance
our knowledge about the radio pulsar and its binary system \cite{kbk96}.
Callanan et al.~\shortcite{cgk98}, for instance, used their results from
optical observations to obtain a number of essential pieces of information for
this system.  Comparing the measured optical luminosity to its value expected
from WD models, they determined the distance to be $d=840\pm$90 pc. The
Doppler shift of the measured H-spectrum of the WD companion gives an
additional radial velocity component of 44$\pm$8 km\,s$^{-1}$ relative to the
solar system barycentre, SSB.  From the radial velocity (semi-)amplitude and
the orbital parameters known from pulsar timing, the mass ratio between the
pulsar and its companion is measured to be 
$m_{\rm PSR}/m_{\rm c}=$10.5$\pm$0.5.

In order to derive the intrinsic spin-down age of millisecond pulsars,
one needs to have accurate distance estimates to correct for apparent
acceleration effects \cite{shk70}.  A proper motion obtained by pulsar
timing is usually converted into a transverse velocity by using a
distance estimate derived from the dispersion in the interstellar
medium, applying a model of the free electron density distribution
\cite{tc93}. However, this model yields a high level of systematic
uncertainty (e.g.~Toscano et al.~1999b\nocite{tbm99}). The dispersion
measure (DM) distance of 520 pc is indeed somewhat smaller than the
distance derived from luminosity models.

Applying classical cooling models \cite{it86} for low mass WDs and the
DM distance, Lorimer et al.  derived an age
of only 0.3~Gyr for the WD companion which was in apparent contradiction with
the inferred spin-down age of the pulsar of 7~Gyr.
However, Alberts et al.\ \shortcite{ash96} were able to model the
binary parameters if they assumed that the cooling age of the WD is
similar to the spin-down age of the pulsar. More recently, Driebe et
al.\ \shortcite{dsb98} modeled the cooling process of WDs, obtaining
an age of 6$\pm $1~Gyr for the companion of PSR~J1012+5307.  Ergma et
al.\ \shortcite{esa98} modeled evolutionary sequences of neutron
stars with close low mass binary companions and short orbital periods,
whilst Sarna et al.\ \shortcite{sam98} developed an evolutionary
scenario for PSR~J1012+5307, reproducing the observed
binary period and companion mass.  They modeled the age of the neutron star
to be between 4.5 and 6.0\,Gyr, whilst the mass of the WD was found to be
$0.16\pm0.02$ M$_\odot$, consistent with mass estimates derived by other
models based on different observations and evolutionary models
\cite{cgk98,dsb98}.


In this paper, we combine a set of high
precision timing data with the information derived for the binary system
from radio pulsar and optical WD observations as well as 
from model calculations.
After a short description of the timing observations and data
reduction techniques, we describe a new binary timing model that is
important for the low eccentricity case, discuss the resulting
solution using statistical tests, and present the derived timing
parameters. These results are analysed in the following sections. We
show that proper motion and Galactic acceleration of PSR~J1012+5307 have a
significant influence on the derived characteristic age. Furthermore,
we obtain strong limits on the true eccentricity of the system and
discuss its meaning for evolutionary scenarios.  As a result of the
stringent limits obtained both on the change in orbital period as well
as on the true eccentricity of the system, we demonstrate that this
binary system is highly suitable for testing different theories of
gravitation.

\section{Observations\label{observe}}

\subsection{Effelsberg timing}

\begin{figure}
\psfig{figure=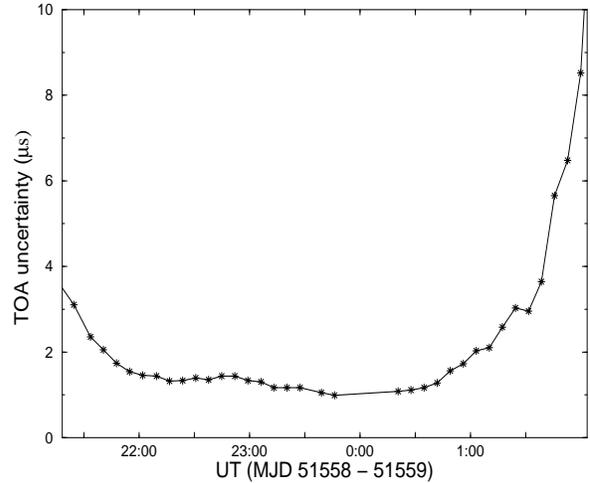,width=8.5cm,height=7cm}
\caption{Errors of the TOAs during a scintillation maximum.\label{1012sci}}
\end{figure}

We have made regular high precision timing observations of
PSR~J1012+5307 since October 1996 using the 100-m radio telescope of the
Max Planck Institut f\"ur Radioastronomie in Effelsberg near Bonn. The
typical observing rate has been once per month with a usual
observing time of about one hour. The overall post-fit RMS of the timing model
applied to the 1213 resulting
times-of-arrival (TOAs) is 3.1 $\mu $s.  Sometimes,
however, the source shows strong intensity fluctuations on time scales
of typically two hours caused by scintillation effects in the ionised
component of the interstellar medium.  We made use of these
intensity maxima by optimising our observing strategy, i.e., staying
``on source'' until the pulsar becomes weaker again. As a consequence
of this highly successful ``scintillation hopping'', many TOAs
are of much higher quality, with errors frequently smaller than
1 $\mu$s. Figure~\ref{1012sci} shows a typical example for
the development of TOA uncertainties during a scintillation maximum.
Since we have many such high quality TOAs covering several orbits,
they proved to be extremely useful for the precise determination of
orbital parameters.

\subsubsection{Receiving systems}

Most of the data were obtained with a 1.3--1.7~GHz tunable HEMT
receiver installed in the primary focus of the telescope. The noise
temperature of this system is 25~K, resulting in a system
temperature from 30~to 50~K on cold sky depending on elevation. 
The antenna gain at these frequencies is 1.5~K~Jy$^{-1}$. 
An intermediate frequency (IF) centred 
on 150 MHz for left-hand (LHC) and right-hand (RHC) 
circularly polarised signals was obtained after down-conversion from
a central RF frequency of usually 1410 MHz.

In order to monitor DM changes, occasionally
we also obtained data at 860 and 2700
MHz.  For the 860-MHz observations we used an uncooled HEMT-receiver
operating in the primary focus with a bandwidth of 30~MHz. The system
temperature during the observations was about 65~K on cold sky
with a telescope gain of about 1.5~K~Jy$^{-1}$.  The 2700-MHz data
were obtained with a cooled HEMT receiver in the secondary focus, with a
system temperature of typically 50~K on cold sky. As for the other
receivers, LHC and RHC signals were mixed down to an IF
of 150~MHz.

\subsubsection{Coherent De-Disperser}

The signals received from the telescope were acquired and processed with the
Effelsberg-Berkeley Pulsar Processor (EBPP) which removes the dispersive
effects of the interstellar medium on-line using the technique of ``coherent
de-dispersion'' \cite{hr75}. Before entering the EBPP, the two IF LHC and RHC
signals are converted to an internal IF of 440 MHz. A maximum bandpass of
112~MHz\footnote{The bandwidth available depends on observing frequency and DM
  of the pulsar. At 1.41~GHz we used a bandwidth of 56 MHz for
  PSR~J1012+5307.} is split into four independent portions for both circular
polarisations. Each portion is mixed down to baseband and subdivided into 8
narrow channels via a set of digital filters \cite{bdz+97} for coherent online
de-dispersion by convolution using a custom VLSI device. 
In total 64 output signals are detected and integrated in phase with the
predicted topocentric pulse period.

\subsubsection{Arrival times measurements}

A TOA was calculated for each average profile
obtained during a 5--10 min observation. During this process,
the observed time-stamped profile was compared to a synthetic template,
which was constructed out of 12 Gaussian components fitted to a 
high signal-to-noise standard profile 
(see Kramer et al.~1998; 1999).
This template matching was done by a least-squares fitting of the
Fourier-transformed data \cite{tay91}.  The final TOA was obtained by using
the measured time delay between the actual profile and the template, and the
accurate time stamp of the data provided by a local H-maser and corrected
off-line to UTC(NIST) using recorded information from the satellites of the
Global Positioning System (GPS).  The uncertainty of each TOA was estimated
using a method described by Downs \& Reichley \shortcite{dr83}.

\subsection{Jodrell Bank timing}

Since its discovery in 1993, PSR~J1012+5307 has been regularly
monitored using the 76-m Lovell telescope at Jodrell Bank with
cryogenic receivers at 408, 606, and 1404\,MHz.  Both LHC and RHC
signals were observed using a $2\times64\times0.125$-MHz filter bank
at 408 and 606\,MHz and a $2\times32\times1.0$-MHz filter bank at
1404\,MHz.  After detection, the signals from the two polarisations
were filtered, digitised at appropriate sampling intervals, and
incoherently de-dispersed in hardware before being folded on-line with
the topocentric pulse period and written to disk.  Each integration
was typically of 1--3 minutes duration; 6 or 12 such integrations
usually constituted an observation.  In the analysis stage, 
the profiles were added
in polarisation pairs before being summed to produce a single
total-intensity profile.  A standard pulse
template was fitted to the observed profiles at each frequency to
determine a total of 400 pulse TOAs. Details of the observing
system and the data reduction scheme can be found elsewhere (e.g.~Bell
et al.~1997\nocite{bbm97}).

Although the templates used in the Effelsberg and Jodrell Bank data
reduction differed, resulting offsets were absorbed in a global
least-squares fit. The typical measurement accuracy of the Jodrell
Bank 1404 MHz timing data is about 9 $\mu$s.

\section{Precision timing}
\label{model}

The combined TOAs, corrected to UTC(NIST) and
weighted by their individual uncertainties
determined in the fitting process, were analysed with the {\sc tempo}
software package\footnote{http://pulsar.princeton.edu/tempo}, 
using the DE200 ephemeris of the Jet
Propulsion Laboratory \cite{sta90}.  Owing to the extremely low
eccentricity of this system, we applied a binary timing model that
uses the Laplace Lagrange parameters $\eta $ and $\kappa $ and
the time of ascending node $T_{\rm ASC}$ in place of the Keplerian
parameters $e$, $\omega $ and $T_0$ (see Appendix~A for details).  The
Keplerian parameters as a function of those used by the model
can be calculated as
\begin{eqnarray}
e       &= & \sqrt{\kappa^2 + \eta^2} \label{ll2kep1} \\
\omega  &= & \arctan (\eta / \kappa )\\
T_0     &= & T_{\rm ASC} + \frac{P_b}{2\pi}\;\arctan (\eta / \kappa ).
\end{eqnarray}

\begin{figure}
\psfig{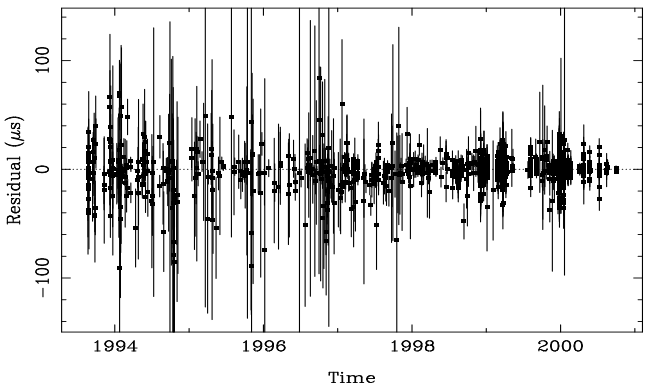}
\caption{Post-fit timing residuals as a function of observing year.\label{reff}}
\end{figure}

\begin{figure}
\psfig{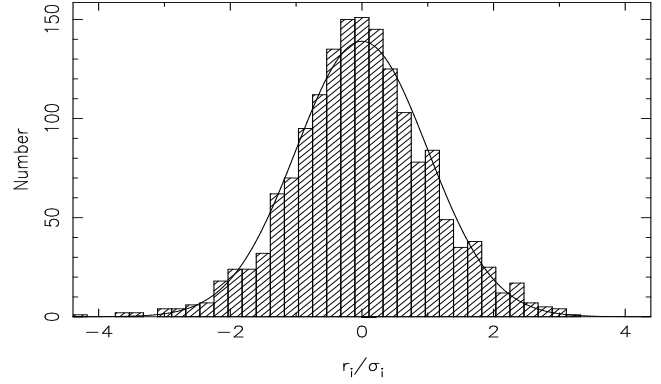}
\caption{Distribution of all post-fit timing residuals normalized by their
uncertainties.\label{geff}}
\end{figure}

{\sc tempo} minimizes the sum of weighted squared timing
residuals, i.e., the difference between observed and model TOAs, 
$r_{\rm i}$, yielding a
set of improved pulsar parameters and post-fit timing residuals.  The
timing residuals shown in Fig.~\ref{reff} indicate deviations from
the timing model which are slightly larger than the calculated TOA
uncertainties, $\sigma_{\rm i}$. 
This most likely stems from a small underestimation of
the error in the TOAs. The data uncertainties used in the following
were scaled by an appropriate factor to achieve a
uniform reduced $\chi^2=1$ for each data set.  In order to test the
statistical properties of our data we have generated histograms of the
deviations of the timing residuals from the model.
The final residuals show the expected Gaussian
distribution (Fig.~\ref{geff}).

\begin{figure}
\psfig{figure=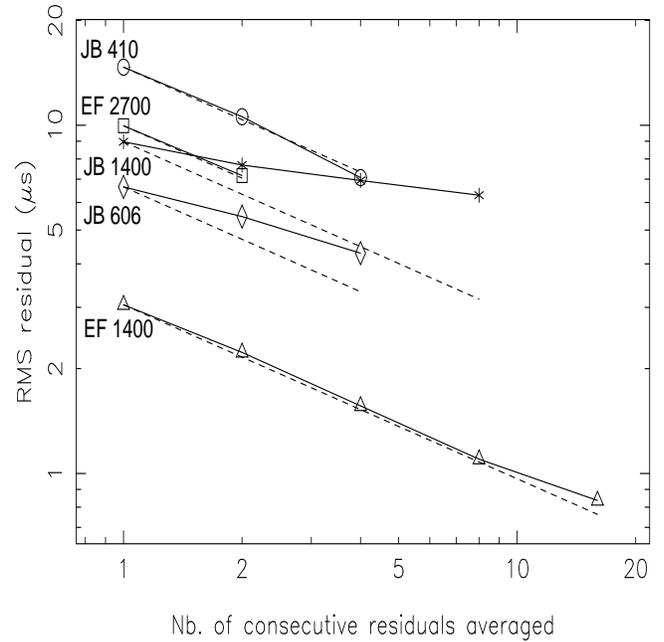,angle=-90,width=8.5cm,height=8.5cm}
\caption{
Mean post-fit RMS residuals of the data sets versus the number
of consecutive averaged timing residuals. The symbols represent
data from:
EBPP 1.4~GHz ($\bigtriangleup$) and 2.7~GHz ($\Box$), and
Jodrell Bank 410~MHz ($\bigcirc$), 606~MHz ($\diamondsuit$), and
1.4~GHz ($\ast$).
The 870~MHz data from Effelsberg are not included here
owing to their small number.\label{rmsPl} }
\end{figure}

In order to search for systematic errors that are either small 
enough to be
individually undetectable or have a time dependence which approximates
some linear combination of terms in the model, we have searched for
tell-tale correlations among nearly adjacent post-fit residuals
(cf.~Taylor \& Weisberg 1989\nocite{tw89}).  
We combined 2, 4, 8 etc.\ consecutive TOAs and evaluated
the average residual of the remaining data set. A
``clean'' data set should exhibit a slope of $-0.5$ in a plot of RMS
versus number of combined TOAs. The results, as shown in
Fig.~\ref{rmsPl}, show no or very little correlation among Effelsberg
data and the Jodrell Bank data recorded at 410 MHz after employing
the timing model described in Table~\ref{params}. The Jodrell Bank
data at 1400 MHz and 606~MHz exhibit, however, some correlation. We
tested for possible impact of these correlations on the fit 
parameters by analysing the
timing data without one or both of these data sub-sets. The parameters
derived from these reduced data sets did not significantly differ from
those derived by the fit to all the data.

The final parameters obtained from fitting the timing model to our data of
PSR~J1012+5307 are shown in Table~\ref{params}. 
The upper limits on the derivatives of
$e$, $x$ and $P_b$ were obtained by separately fitting for each of them. These
were set to zero during the fits for all other parameters.  
The residuals show slow variations that can be removed by fitting
for $\ddot{P}$. 
We do not believe that instrumental effects can account
for this effect but rather suggest that the measurement
 reflects rotational instabilities
in the pulsar. In this case, PSR~J1012+5307 is only the second Galactic field
millisecond pulsar to show a significant second spin frequency derivative,
$\ddot \nu$, the other being PSR~B1937+21 \cite{ktr94}\footnote{Cognard et
  al.~\shortcite{cbl+96} measured a $\ddot \nu$ for the millisecond pulsar
  B1821$-$24 located in a globular cluster.  Here the
  acceleration by the cluster gravitational potential may significantly
  affect the observed spin-down parameters.}. We will discuss the 
possible important implications of this result in Section \ref{further}.

\begin{table}
\begin{tabular}{ll}
\hline
\\
Right Ascension, $\alpha $ (J2000)      & $10^{\rm h} 12^{\rm m} 33\fs 43370(3)$\\
Declination,     $\delta $ (J2000)      & $53^\circ 07\arcmin 02\farcs 5884(4)$\\
$\mu_\alpha $ (mas yr$^{-1}$)           & $2.4(2)$\\
$\mu_\delta $ (mas yr$^{-1}$)           & $-25.2(2)$\\
\\
$\nu $ (Hz)                             & $190.267837621903(4)$\\
$\dot \nu $ (s$^{-2}$)                  & $-6.2029(4)\cdot 10^{-16}$\\
$\ddot \nu $ (s$^{-3}$)                 & $-9.8(2.1)\cdot 10^{-27}$\\
\\
$P $ (ms)                               & $5.2557490141197(1)$\\
$\dot P $                               & $1.7134(1)\cdot 10^{-20}$\\
$\ddot P $ (s$^{-1}$)                   & $5.1(1.1)\cdot 10^{-31}$\\
\\
Epoch (MJD)                             & $50700.0$\\
\\
Dispersion Measure (DM) (cm$^{-3}$ pc )     & $9.0233(2)$\\
\hline
\\
Orbital period, $P_b$ (days)            & $0.60467271355(8)$\\
Proj.\ semi-major axis, $x$ (s)         & $0.5818172(2)$\\
$\eta$                                  & $0.9(8)\cdot 10^{-6}$\\
$\kappa$                                & $0.01(80)\cdot 10^{-6}$\\
$T_{\rm ASC}$ (MJD)                     & $50700.0816290(1)$\\
\\
\hline
\\
Upper limits: & \\
\\
Parallax, $\pi $(mas)                 & $<\,1.3$\\
$e$                                     & $<1.3\times 10^{-6}$ $^a$ \\
$\dot e$ (s$^{-1}$)                     & $<2\cdot 10^{-14}$\\
$\dot x$                                & $<1.4\cdot 10^{-14}$\\
$\dot P_b$                              & $<1\cdot 10^{-13}$\\
$\dot{\rm DM}$ (cm$^{-3}$ pc yr$^{-1}$) & $<1.2\cdot 10^{-4}$ \\
\hline
\end{tabular}

{\small $^a$ See section \ref{trueEcc} for details.}

\caption{
Timing parameters for the millisecond pulsar J1012+5307. Quoted errors
correspond to twice the errors derived by {\sc tempo}. Upper limits
represent 95\% C.L.
\label{params}}
\end{table}

\section{Intrinsic Pulsar Parameters}


As summarised in Section \ref{intro} the binary millisecond pulsar
PSR~J1012+5307 is unique since the combination of optical and radio
measurements allows a full determination of the motion of the system
through space and also the determination of several intrinsic parameters.


\subsection{3-D velocity}

The full 3-d motion of the pulsar relative to the SSB is determined by
combining the timing proper motion with distance to the system obtained from
luminosity models for the white dwarf companion, $d=840\pm90$ pc.  This
distance is consistent with the obtained upper limit for the timing parallax.
Given the pulsar's ecliptic latitude of 38$^\circ$, 
it is not unexpected that a
significant value for the timing parallax could not be obtained. We derive 
transverse velocities of
\begin{eqnarray}
v_\alpha = \mu_\alpha \cdot d &=& 8.8\pm1.2\;{\rm km\,s}^{-1}\\
v_\delta = \mu_\delta \cdot d &=& -102\pm11\;{\rm km\,s}^{-1}.
\end{eqnarray}
Optical measurements (cf. Section~\ref{intro}) yield a radial velocity
component of +44(8) km\,s$^{-1}$.  Hence, the total velocity of the
system is 111(28) km\,s$^{-1}$.  This value is consistent
with the average space velocity of millisecond pulsars of 130 km s$^{-1}$
estimated by Lyne et al.~\shortcite{lml+98}
and Toscano et al.\ \shortcite{tsb99}.

\subsection{Doppler effects}

Due to the motion of PSR~J1012+5307 relative to the solar system, 
any period $P^{\rm obs}$ of the pulsar spin or binary system, measured
at the SSB, differs from the period $P^{\rm int}$ in the pulsar
reference frame by a Doppler factor $D$. Following Damour \& Taylor
\shortcite{dt91}, we write the relation between these periods
\begin{equation}\label{P-intrinsic}
P^{\rm obs}
= D\,P^{\rm intr}
\equiv 
(1+\frac{\bf n\, \cdot \, v_r}{c})\,P^{\rm intr}
+ {\cal O}(v_r^2/c^2)
\end{equation}
with ${\bf n}$ the unit vector to the pulsar and ${\bf v_r}$ the relative
velocity between the pulsar and the SSB \cite{dt92}. As
$v_r/c$ is typically less than 0.1\% for millisecond pulsars,
this Doppler shift
is for most purposes unimportant. Hence, we will for the rest of this
paper denote $P^{\rm obs}$ as $P$. However, the period derivative may be
modified by the relative motion.

Calculating the time-derivative of Eqn.~\ref{P-intrinsic} and
separating the effect of the proper motion of the system \cite{shk70}
from the influence of the Galactic acceleration, we derive as
contributions to the period derivative
\begin{equation}
\dot P^{\rm obs}
= \dot P^{\rm intr}\, D + \dot P^{\rm Shk} + \dot P^{\rm Gal},
\end{equation}
where the difference of the constant Doppler factor $D$ from unity 
may be neglected. The
contributions $\dot P^{\rm Shk} + \dot P^{\rm Gal} \equiv \dot P^{\rm
D}$ are written explicitly
\begin{eqnarray}
\dot P^{\rm Shk} & = & \frac{1}{c}\;\mu^2 \,d\,P \label{shkInt}\\
\dot P^{\rm Gal} & = & \frac{1}{c}\;\left(
        {\bf n_{\rm psr}\cdot } [{\bf a_{\rm psr} - a_{\rm e}}]
\right)\, P\label{galInt},
\end{eqnarray}
where $d$ is the distance to the pulsar. Here ${\bf n_{\rm
psr}}$ is the position vector to the pulsar, $\mu $ its proper motion
and ${\bf a_{\rm psr}}$ and ${\bf a_{\rm e}}$ the Galactic
acceleration of the pulsar and the Earth, respectively.
We measure ${\bf n_{\rm
psr}}$ and $\mu $ from our timing observations. The small correction
term $\dot P^{\rm Gal}/P = -7\times 10^{-20} $s$^{-1}$ for the Galactic
acceleration is obtained applying a model for the Galactic potential
\cite{ci87,kg89}. Inserting these parameters into Eqns.
\ref{shkInt} and \ref{galInt}, we obtain a Doppler correction of
\begin{eqnarray}\label{dotPd1012}
\dot P^{D}/P       &=& \ \ 1.4\pm0.3\times 10^{-18}\,{\rm s}^{-1}.
\end{eqnarray}
Its error is dominated by the
uncertainty in the distance to the pulsar.

\subsection{Characteristic age}

We can now use our timing results to obtain the true spin-down age,
the so-called characteristic age of PSR~J1012+5307. 
With a spin period of $P=5.256\,$ms and a
Doppler factor as given in Eqn.~\ref{dotPd1012}, we derive
a Doppler correction of 
\begin{equation}
\dot P^{D}=7.4\pm1.6\times 10^{-21},
\end{equation}
which has to be subtracted from the measured $\dot P$ before calculating
the intrinsic characteristic age of the system:
\begin{equation}
t_{\rm char}^{\rm intr}=\frac{P}{2\,\dot P^{\rm int}}
=\frac{P}{2\,(\dot P^{\rm obs} - \dot P^{\rm D})} = 8.6\pm1.9\,{\rm Gyr.}
\end{equation}
We note that 
the characteristic age assumes magnetic dipole braking (i.e., braking 
index of 3) and is only a realistic estimate of the real pulsar age 
if the initial spin period was sufficiently small (Camilo
et al.~1994\nocite{ctk94}). Although this may not apply to He-WD binaries
(Backer 1998\nocite{b98}), the characteristic age 
is in reasonable agreement with a cooling age of the white dwarf as 
derived by evolutionary models \cite{ash96,sam98,dsb98}.

\subsection{Orbital eccentricity}

\label{trueEcc}
In Appendix~A we analyse the contribution of the Shapiro
delay for binary orbits with moderate inclination angles and
negligible intrinsic eccentricities to the apparent eccentricity of
the system. As we show, the Shapiro delay cannot be separated from the
Roemer delay, which leads to a small correction to the observed
eccentricity.  In order to investigate this effect for PSR~J1012+5307
quantitatively and to obtain the true orbital eccentricity, we make use 
of independent determinations of the system parameters.

\begin{figure}
\psfig{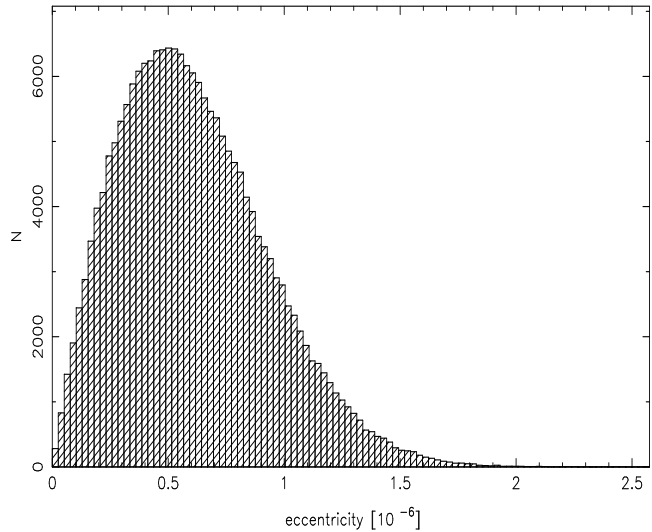}
\caption{Distribution of values for the real eccentricity from simulated
parameters (see text).\label{1012ecc}}
\end{figure}

From the companion mass of 0.16(2) M$_\odot$, the measured mass ratio
of $q\equiv m_p/m_c=10.5(5)$, and the mass function
\begin{equation}
f_{\rm m}\equiv \frac{m_c^3 \sin^3 i}{\left(m_c + m_p\right)^2}
=0.000578\;{\rm M}_\odot
\end{equation}
obtained from the timing analysis, we derive the range $r$ and shape
$s$ of the Shapiro delay in the system according to
\begin{eqnarray}
r[\mu {\rm s}] & = & 4.9255\;(m_c/{\rm M}_\odot) \qquad \mbox{\rm and}\\
s  \equiv \sin i & = & \left( \frac{f_{\rm m}\,(q+1)^2}{m_c} \right)^{1/3}.
\end{eqnarray}
The resulting shape parameter of $s\approx 0.8$ indicates a moderate
inclination of the system, i.e., $i=52^\circ$ or $128^\circ$.  The
contribution of the Shapiro delay to the observed $\eta$ is calculated
according to Eqn.~\ref{etaobs} 
and subtracted from the measured value.
Applying Eqn.~\ref{ll2kep1}, we obtain upper limits on
the true eccentricity of
\begin{equation}
\begin{array}{l}
   \displaystyle
   e^{\rm intr}<0.8\times 10^{-6}~~~~~\mbox{(68\% C.L.)} \\[2mm]
   \displaystyle
   e^{\rm intr}<1.3\times 10^{-6}~~~~~\mbox{(95\% C.L.)},
\end{array}
\end{equation}
which were derived from Monte Carlo simulations. In these
calculations, whose results are displayed in Fig.~\ref{1012ecc}, we
simulated the parameters $q$ and $m_c$ and the observed $\kappa$ and
$\eta$ in accordance with the observational uncertainties.

\section{Tests of theories of gravitation}

\subsection{Dipole gravitational waves\label{dipole_grav}} 

Unlike general relativity, many alternative theories of gravity
predict the presence of a radiative monopole and dipole as well as the
quadrupole and higher order multipoles \cite{wil93}. Scalar-tensor
theories, for instance, predict for binary systems a loss of orbital
energy which is at highest order dominated by scalar dipole
radiation. As a result, the period $P_b$ of a circular orbit will
change with
\begin{equation}\label{gdr1}
   \dot P^{{\rm dipole}}_{b} \simeq -\frac{4\pi^2 G_*}{c^3\,P_b}
        \frac{m_p\;m_c}{m_p + m_c} \; (\alpha_{\rm p} - \alpha_{\rm
        c})^2,
\end{equation}
where $G_*$ is the bare gravitational constant, $c$ is the speed of
light and $m_p$ and $m_c$ the mass of the pulsar and its companion
\cite{def96}. The parameters $\alpha_{\rm p}$ and $\alpha_{\rm c}$
represent the effective coupling strengths of the scalar field to the
pulsar and its companion, respectively.

Damour \& Esposito-Far\`ese \shortcite{def96} show that $\alpha_{\rm c}$ can
be neglected compared to $\alpha_{\rm p}$, since the gravitational binding
energy per unit mass of a WD, $\varepsilon_{\rm WD} \equiv E^{\rm grav}/mc^2
\approx 10^{-4}$, is very much smaller than that of a 1.4-M$_\odot$ neutron
star, $\varepsilon_{\rm p}\approx 0.15$.

Using this assumption, and taking $G_*\approx G$, Eqn.~\ref{gdr1}
can be approximated as:
\begin{equation}\label{gdr2}
   \dot P^{{\rm dipole}}_{b} \simeq -\frac{4\pi^2 G}{c^3\,P_b}\;
        m_c\;\frac{q}{q+1} \; \alpha_{\rm p}^2.
\end{equation}
Since all terms except $\alpha_{\rm p}$ in this equation are measured,
the effective coupling strength of PSR~J1012+5307 can be restricted by
limits on $\dot P_b$. We will show that these are the tightest bounds
on $\alpha_{\rm p}^2 $ ever measured for a neutron star.

The observable rate of change of the orbital period of
PSR~J1012+5307 is the sum of several contributions, either 
intrinsic to the orbit
or caused by projection effects:
\begin{equation}
\dot P_b
= \dot P_b^{{\rm D}} 
+ \dot P_b^{{\rm GR}} 
+ \dot P_b^{{\rm \dot G}} 
+ \dot P^{{\rm dipole}}_{b}
\end{equation}
\cite{dt91}. Here, the Doppler correction of $\dot P_b$ is given by
$\dot P_b^{{\rm D}} $ and the contribution $\dot P_b^{{\rm GR}}$
represents gravitational radiation as predicted by general
relativity. The values of these effects can be estimated and may be
subtracted from the measured value of $\dot P_b$. The additional
contributions $\dot P_b^{{\rm \dot G}}$ and $\dot P^{\rm dipole}_{b}$
represent the period derivative due to a change of the gravitation
constant and due to gravitational dipole radiation, respectively. All
terms of gravitational radiation not accounted for are of the order
$c^{-5}$ or higher.
%
We derive $\dot P_b^{\rm D}$ from Eqn.~\ref{dotPd1012}, with the orbital
period of 52\,240\,s, to be
\begin{eqnarray}
\dot P_b^{\rm D } = 7.3\pm1.6\times 10^{-14}.
\end{eqnarray}
The orbital period decay derived from the theory of general
relativity for a circular orbit can be approximated to
\begin{equation}
   \dot P_b^{{\rm GR}}
\simeq -\frac{384\,\pi^2}{5} \; 
\frac{\left( G\,m_c\right)^{5/3} }{P_b^{5/3}c^5} \; \frac{q}{(q+1)^{1/3}}
\approx - 1.0\pm0.2\times 10^{-14},
\end{equation}
using the WD-mass $m_c=0.16(2)$ M$_\odot$ and the mass ratio of
$q\approx 10.5$ between the pulsar and the WD \cite{cgk98}.

A variation in the orbital period as a result of a possible time
dependence of the gravitational constant can for this system be
approximated to
\begin{equation}
\dot P_b^{{\rm \dot G}}
  \approx -2\,\frac{\dot G}{G}(1 + c_p)\,P_b
\end{equation}
\cite{nor90} where $c_p$ is the compactness of the pulsar, which is
approximately two times its binding energy. Using the upper limit
$|\dot G/G| \la 8\times 10^{-12} {\rm yr}^{-1}$ \cite{wnd96}
we can restrict
\begin{equation}
|\dot P_b^{{\rm \dot G}}| \la 3.6\times 10^{-14}.
\end{equation}
This value is only about 1/3 of our 1-$\sigma $ error for $\dot
P_b$. We can therefore exclude a significant influence on $\dot P_b$
from this effect.

Subtracting the known contributions from the limit measured for $\dot P_b$
yields a value of $\dot P^{{\rm dipole}}_{b} \approx -0.6(1.1)\,10^{-13}$. 
Using this value we obtain limits for $\alpha_{\rm p}$ of
\begin{equation}
\begin{array}{l}
   \displaystyle
   \alpha_{\rm p}^2 < 2\times 10^{-4}~~~~~\mbox{(68\% C.L.)} \\[2mm]
   \displaystyle
   \alpha_{\rm p}^2 < 4\times 10^{-4}~~~~~\mbox{(95\% C.L.)},
\end{array}
\end{equation}
where we have taken into account uncertainties in the parameter
values used. 

Simulations of future observations show that the precision of
the measurement of $\dot P_b$ will increase significantly.
A continuation of our timing observations with an accuracy and
observing rate similar to that of the past few years will lead to an
improvement by a factor of about 3 within the next three years. The
applicability of this improvement for the value of $\alpha_{\rm p} $
might be reduced by the unknown contribution of $\dot G$, unless the
limits on this parameter are also improved. It is also desirable
that the distance to PSR~J1012+5307 be determined with higher accuracy to
allow a more reliable estimate of the transverse Doppler effect.

\subsection{Local Lorentz invariance} 

Will \& Nordtvedt (1972) pointed out that one expects the existence of a
preferred cosmic rest frame for gravitational interaction if gravity is
mediated in part by a long range vector field or by a second tensor field. In
the post-Newtonian limit all the gravitational effects associated with such a
preferred frame are phenomenologically describable by two parameters,
$\alpha_1$ and $\alpha_2$. Only in gravity theories without a preferred frame,
e.g.\ general relativity, are the values of these parameters equal to zero.

A consequence of a preferred cosmic rest frame would be
the introduction of an eccentricity 
into orbits of gravitationally bound bodies of different
masses which are in motion relative to this preferred frame. This effect
is quantified by the parameter $\alpha_1$. Wex \shortcite{wex00} has
applied a statistical analysis of all pulsar binary systems to find
limits for this parameter. This analysis yields $|\alpha_1|
<1.4\times 10^{-4}$ at 95\% C.L. Due to its extremely low true
eccentricity and the known three dimensional proper motion,
the data presented here for PSR~J1012+5307 and included in Wex's analysis 
play a central role in this analysis. They might provide
more stringent tests of local Lorentz invariance, if the real
eccentricity of the system is even smaller and better observational
limits on it are found.




\section{Discussion}
\label{further}

\subsection{Timing noise}

Over the observed time span of about 7 years, PSR~J1012+5307 shows a
significant deviation from a simple $\nu $--$\dot \nu $ spin down
behaviour. In order to model this effect, the spin-down model can be
extended by allowing for a non-zero second derivative of
the spin frequency, $\ddot \nu$ (e.g.~Cordes \& Downes 1985). However,
rather than being the result of rotational irregularities of the
neutron star itself, a possible alternative explanation for a second
spin frequency derivative could be a third order term of the proper
motion,
\begin{equation}
\ddot \nu^{\rm D} =
        - 3\;\frac{v_r}{c} \mu^2\;\nu_{\rm intr}
        - 2\;\frac{d\,\mu^2 }{c}\; \dot \nu_{\rm intr},
\end{equation}
where $v_r$ is the radial velocity of the system \cite{phi92}. This
term can be computed and it is of order 10$^{-30}\, $s$^{-3}$,
i.e., too small to explain the measured effect. 

In order to investigate
the possibility that the observed timing noise is caused by small
changes in the DM, we modeled our multi-frequency
 data with a simultaneous fit of
$\ddot \nu $ and a second-order polynomial for DM. While $\dot
{\rm DM}$ and $\ddot {\rm DM}$ were not significant, the $\ddot \nu$ measurement
was still a 5-$\sigma$ effect. The upper limit on $\dot{\rm DM}$ given
in Table~\ref{params} is interestingly much smaller than one would expect
for DM=9.02 cm$^{-3}$ pc (Backer et al.~1993\nocite{bhvf93}).

As a possible explanation, a second period derivative in the timing
residuals can also be the result of a second light companion in a wider
orbit \cite{tac99}. Assuming a circular orbit and an intermediate
orbital phase, the measured value for $\ddot \nu$ can be explained
with e.g.\ a planet of terrestrial mass at $\sim30$~AU distance or a
Jupiter-like planet at $\sim170$~AU distance from the centre of
mass. Future observations will show if the timing residuals can be
explained by a second Keplerian orbit.

However, it is more likely that the irregularities are
intrinsic to the pulsar itself. Timing noise has been studied for
normal, slowly rotating pulsars. Arzoumanian et al.\
\shortcite{ant94}, for instance, defined the stability parameter
\begin{equation}
\Delta(t) \equiv \log \left( \frac{|\ddot \nu |}{6\,\nu}\;t^3
        \right),
\end{equation}
where $t$ is the observing span which in their case was $10^8$ s.
Despite noticeable scatter, they obtained a correlation of this
parameter with the period derivative for slowly rotating pulsars.  
This correlation predicts a value of $\Delta_8 \approx -5.3$ for
PSR~J1012+5307. Using timing data over a time span $10^8$ s, we
measure a value of $\Delta_8 \approx -5$. Given that the fit leading to
Arzoumanian et al.'s relation was made by eye, both values are in
excellent agreement, in particular given the large scatter seen for
normal pulsars. Moreover, the value of the second spin frequency
derivative, $\ddot \nu = -9.8(2.1) \times 10^{-27}$s$^{-3}$, is similar
to the value of $\ddot \nu$ published for
PSR~B1937+21 \cite{ktr94}. Both sources exhibit a $\Delta_8$ consistent
with the correlation derived by Arzoumanian et al.  Although we cannot 
yet completely rule out a different origin of the measured period
second derivative, the value determined is apparently
consistent with what is known for timing noise 
from normal pulsars. Assuming that we are now indeed observing
rotational irregularities in two Galactic field
millisecond pulsars, being consistent with
the behaviour of normal pulsars extrapolated to very small period
derivatives, it may indicate that the empirical relation found by 
Arzoumanian et al.\ is generally
valid for millisecond pulsars as well as for normal pulsars. 
As
other millisecond pulsars are now being monitored with similar timing
precision over similar time spans, this conjecture can be tested.
It may well be the case that we are finally exploring
the ultimate accuracy of the pulsar clockwork, deciding over the
suitability of millisecond pulsars as potential standards of time.

\subsection{Orbital eccentricity --- a relic of binary evolution}

\begin{figure}
\psfig{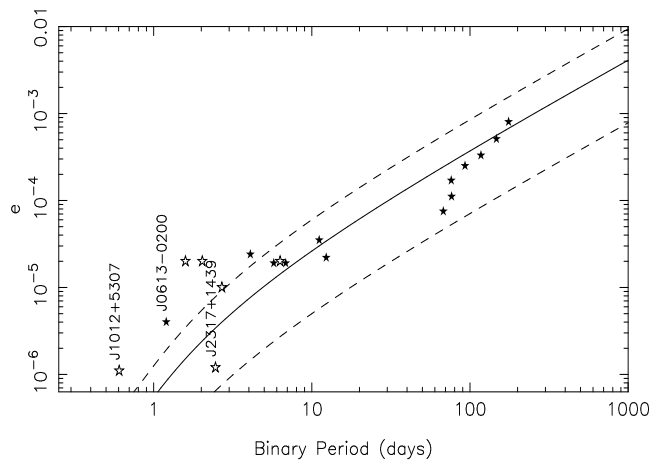}
\caption{Relation between binary period and orbital eccentricity 
(filled symbols are measurements, open symbols upper limits) in binary
systems with a millisecond pulsar and white dwarf companion.\label{eccplt}}
\end{figure}

The convective fluctuation-dissipation theory of Phinney
\shortcite{phi92} predicts a strong correlation between the orbital
eccentricities and orbital periods of binary pulsars that have been
recycled by stable mass-transfer from a Roche-lobe-filling, low-mass
red giant. The measurements and upper limits for the orbital
eccentricities of systems with appropriate evolutionary history are
plotted in Fig.\,\ref{eccplt}. They are, with the exception of
PSR~J0613$-$0200, in excellent agreement with the model of Phinney \&
Kulkarni (1994; curves in Fig.~\ref{eccplt}).\nocite{pk94}

The lowest orbital eccentricity of a millisecond binary pulsar system
known is the 
upper limit of 1.3$\times $10$^{-6}$ (68\% C.L.) derived by Camilo et
al.\ \shortcite{cnt96} for PSR~J2317+1439, which has an orbital period
of 2.5 days. While this source follows the predicted trend, the
eccentricity of the 1.2-d binary pulsar J0613$-$0200 has been
determined to be $3.8\pm1.0\times 10^{-6}$, which exceeds the
95\% C.L. of this model relation by a factor of about 2.5
(Fig.~\ref{eccplt}). However, Phinney \& Kulkarni \shortcite{pk94}
pointed out that for extremely close binary systems the orbital
eccentricity is difficult to model. Therefore, it is interesting
to note that the upper limit for the eccentricity of the 0.6-d binary
pulsar J1012+5307 of only $e<0.8\times10^{-6}$ (68\%
C.L.) is significantly
lower than that of PSR~J0613$-$0200.

\subsection{Future observations}

The motion of the binary system relative to the SSB results in a
change of the orbital inclination that causes a variation in the
projected semi-major axis of the system 
\cite{ajr96,kop96,bbm97,sbm97}. The amplitude of this effect is given by
\begin{equation}
\dot x_\mu      = 1.54\times 10^{-16}x\,\cot i\,
(-\mu_\alpha\,\sin \Omega\,+\mu_\delta\,\cos \Omega)
\end{equation}
where $\Omega $ is the position angle of the ascending node and
$\mu_\alpha $ and $\mu_\delta $ the components of the proper motion
in mas$\,$yr$^{-1}$. Given the value of the orbital inclination
$\sin i\approx 0.8$ as determined from the white dwarf optical light curve, 
a measurement of $\dot x$ allows one to restrict the
orbital orientation $\Omega $.
%
Using our current data set, we derive a 1-$\sigma $ uncertainty for $\dot x$
that is about twice the value expected for a perfect alignment between
$\Omega $ and $\mu $.  Continuing timing observation will 
permit the measurement of this effect.  Simulations, assuming an
accuracy and data-taking rate similar to those at present,
show that a 3-$\sigma$ detection of $\dot x$ should be possible
within the next four or five years. This would finally complete 
our knowledge of the full orientation and motion of the PSR~J1012+5307
system.

\section{Summary}

We have performed long-term high-precision timing observations of
PSR~J1012+5307. Combining our measurements with results based on optical
observations, we derived complete 3-D velocity information for this
system, permitting the correction of the measured spin parameters for Doppler
effects.  Due to the precision of our measurements and the 
extremely small eccentricity of this binary pulsar, we could use it as
an ideal laboratory to test certain aspects of theories of
gravitation.  The deviation of the timing data from a simple spin-down
model is probably due to rotational instabilities which are 
consistent with the extrapolation of the known spin-down behaviour of
slowly rotating pulsars, suggesting that this phenomenon is a common
property of most pulsars.

\section*{Acknowledgements}

We are grateful to all staff at the Effelsberg and Jodrell Bank
observatories for their help with the observations.  F.C. is
supported by NASA grant NAG~5-9095. O.D.~acknowledges the receipt 
of an Alexander-von-Humboldt fellowship.

\appendix
\label{appendix}

\section{Timing of small-eccentricity binary pulsars}

Neglecting relativistic effects and parallax, the barycentric arrival time,
$t_b$, of a signal emitted by a pulsar in an orbit around a compact companion
is given by
\begin{equation}\label{TF3}
   t_b-t_0=T+\Delta_R(T).
\end{equation}
$T$ denotes the time of emission of the pulse signal as measured in
the reference frame of the pulsar, i.e., it is directly related to the
rotational phase $\phi$ of the pulsar by
\begin{equation}
   \phi \equiv \phi_0 + \nu T + \frac{1}{2}\dot\nu T^2 + \dots
\end{equation}
The Roemer delay, $\Delta_R$, caused by the orbital motion of the
pulsar is given by \cite{bt76}
\begin{equation} \label{roemer}
   \Delta_R = x\left[(\cos U-e)\;\sin\omega+
               \sin U(1-e^2)^{1/2}\;\cos\omega\right].
\end{equation}
where the eccentric anomaly, $U$, is determined by Kepler's equation
\begin{equation}
   U-e\sin U = n_b\;(T-T_0)\;, \quad n_b \equiv 2\pi/P_b,
\end{equation}
and $\omega$ denotes the longitude of periastron.

For small-eccentricity binary pulsars, however, the location of
periastron is not a prominent feature in the TOAs.  Therefore,
modeling the timing data of small-eccentricity binary pulsars with
equation (\ref{roemer}) leads to very high correlations between the
parameters $\omega$ and $T_0$. As a result, $\omega$ and $T_0$
show unacceptably large uncertainties in the $\chi^2$ estimation of
these parameters.

\subsection{A timing model for small-eccentricity binary pulsars}

Neglecting terms of order $e^2$ the orbital motion ${\bf X}=R(\cos
\varphi, \sin \varphi,0)$ of a pulsar in a small-eccentricity binary
system, is given by (see, e.g., Roy~1988\nocite{Roy88})
\begin{equation}
\left.\begin{array}{lcl}
   R &=& a_p\;(1-e\cos M) \\[3mm] \varphi &=& M+2e\;\sin M
\end{array}\right\} \quad  M=n_b\;(T-T_0),
\end{equation}
where $M = \varphi = 0$ corresponds to the location of periastron.
To first order approximation in the small eccentricity $e$ the Roemer
delay (Eq.~\ref{roemer}) can therefore be written as
\begin{equation}\label{roemer2}
\begin{array}{l}
   \displaystyle   
   \Delta_R \simeq  x\left( \sin\Phi+\frac{\kappa}{2}\sin2\Phi
                    -\frac{\eta}{2}\cos2\Phi\right), \\[3mm]
   \displaystyle
   \Phi = n_b\;(T-T_{{\rm asc}}),
\end{array}
\end{equation}
where terms which are constant in time are omitted.  The three
Keplerian parameters $T_0$, $e$, and $\omega$ are replaced by the
{\it time of ascending node\/} which is defined by
\begin{equation}
   T_{{\rm asc}} \equiv T_0 - \omega/n_b,
\end{equation}
and the {\it first\/} and {\it second Laplace-Lagrange parameter\/},
\begin{equation}
   \eta   \equiv e\;\sin\omega \qquad \mbox{and} \qquad
   \kappa \equiv e\;\cos\omega,
\end{equation}
respectively. Note that the actual time when the pulsar passes through the
ascending node is given by $T_{{\rm asc}}+2\eta/n_b$. The time of
conjunction is given by $T_{{\rm asc}}+P_b/4-2\kappa/n_b$.

Eq.~(\ref{roemer2}) accounts only for first-order corrections in $e$.
Therefore, the difference between the exact expression (\ref{roemer})
and Eq.~(\ref{roemer2}) can grow up to $xe^2$.  For most of the low
eccentricity binary pulsars, the error in the TOA measurements is much
larger than $xe^2$ and thus the linear-in-$e$ model is sufficient.

Secular changes in the parameters $n_b$, $x$, $e$, and $\omega$ in
Eq.~(\ref{roemer}) are accounted for in the new timing model by using
\begin{equation} \label{secch1}
   \Phi = \bar{n}_b\;(T-T_{{\rm asc}})
            +\frac{1}{2}\;\dot{\bar{n}}_b\;(T-T_{{\rm asc}}),
\end{equation}
and
\begin{equation}
\begin{array}{lcl}
   x      &=& x_0      + \dot x    \;(T-T_{{\rm asc}}),\\
   \eta   &=& \eta_0   + \dot\eta  \;(T-T_{{\rm asc}}),\\
   \kappa &=& \kappa_0 + \dot\kappa\;(T-T_{{\rm asc}}),
   \label{secch2}
\end{array}
\end{equation}
in Eq.~(\ref{roemer2}) where the relations
\begin{eqnarray}
   \bar n_b        &=& n_b + \dot\omega - \dot n_b(T_0-T_{{\rm
asc}}), \\
   T_{{\rm asc}}   &=& T_0-\frac{\omega_0}{n_b+\dot\omega}, \\
   \dot{\bar{n}}_b &=& \dot n_b, \\
   \dot\eta        &=& \dot e\;\sin\omega + e\;\cos\omega\;\dot\omega, \\
   \dot\kappa      &=& \dot e\;\cos\omega - e\;\sin\omega\;\dot\omega 
\end{eqnarray}
hold.

\subsection{Small-eccentricity orbits and Shapiro delay}

For small-eccentricity binary pulsars, the Shapiro delay can be written as
\begin{equation} \label{shap}   
   \Delta_S = -2r\;\ln(1-s \sin\Phi),
\end{equation}
where $r=Gm_c/c^3$ and $s=\sin i$. As a Fourier series
Eq.~(\ref{shap}) takes the form
\begin{equation} \label{shap2}
   \Delta_S = 2r(a_0 + b_1 \sin\Phi - a_2 \cos 2\Phi + \dots),
\end{equation}
where 
\begin{eqnarray}
   a_0 &=& -\ln\left(\frac{1+\sqrt{1-s^2}}{2}\right), \\
   b_1 &=& 2\frac{1-\sqrt{1-s^2}}{s}, \\
   a_2 &=& 2\frac{1-\sqrt{1-s^2}}{s^2} - 1.
\end{eqnarray}
Only for orbits where $\sqrt{1-s^2} \ll 1$ (nearly edge-on) are higher
harmonics, indicated as $\dots$ in Eq.~(\ref{shap2}),
significant. Otherwise, the Shapiro delay cannot be separated from the
Roemer delay. Consequently the observed values for $x$ and $\eta$
differ from their intrinsic values by
\begin{eqnarray}
   x^{(obs)}     &=& x + 2rb_1, \\
   \eta^{(obs)}  &=& \eta + 4ra_2/x \label{etaobs},
\end{eqnarray}
as can be seen by a comparison of Eq.~(\ref{shap2}) with
Eq.~(\ref{roemer2}).  In terms of the Blandford-Teukolsky model, Eq.~(\ref{etaobs})
means that the observed value of the eccentricity $e$ and the observed
value of the longitude of periastron $\omega$ are different from the
intrinsic values of these parameters.

This new model has been implemented in the {\sc tempo} timing software
as binary model ELL1.

\label{lastpage}

\end{document}